%% file: main.tex
\documentclass[10pt,conference]{IEEEtran}

\newcommand{\keywords}[1]{\textbf{Keywords:} #1}

\usepackage{amsmath,amssymb,amsfonts}
\usepackage{nicefrac}
\usepackage{braket}

\usepackage{graphicx}
\usepackage{textcomp}
\usepackage{booktabs}
\usepackage{multirow}
\usepackage{makecell}
\usepackage{orcidlink}
\usepackage{adjustbox}
\usepackage{microtype}
\usepackage{setspace}
\usepackage{enumitem}
\setlist[itemize]{leftmargin=*}
\setlist[enumerate]{leftmargin=*}

\usepackage{xcolor}
\usepackage[most]{tcolorbox}
\tcbuselibrary{listings,skins,breakable}

\usepackage{listings}

\usepackage{algorithm}
\usepackage{algpseudocode}

\usepackage{tikz}
\usetikzlibrary{arrows.meta,positioning,fit,calc}

\usepackage{orcidlink}
\usepackage{comment}
\usepackage{url}

\usepackage{hyperref}
\usepackage{cleveref}

\newtcolorbox[auto counter, number within=subsection]{mybox}[2][]{%
  enhanced,
  colback=black!3!white,
  colframe=black!90!black,
  breakable,
  left=0.5pt,
  right=0.5pt,
  top=0.5pt,
  bottom=0.5pt,
  boxsep=0.5pt,
  boxrule=0.5pt,
  title=Box~\thetcbcounter\ -- #2,
  label type=mybox,
  #1
}

\crefname{mybox}{Box}{Boxes}
\Crefname{mybox}{Box}{Boxes}

\def\BibTeX{{\rm B\kern-.05em{\sc i\kern-.025em b}\kern-.08em
    T\kern-.1667em\lower.7ex\hbox{E}\kern-.125emX}}

\usepackage{array,makecell}
\usepackage{tabularx}
\newcolumntype{C}[1]{>{\centering\arraybackslash}m{#1}}

\usepackage[font=footnotesize,skip=1.5pt]{caption}
\usepackage{float}

\begin{document}

\title{Q-RAIL: A Reliability-Aware Framework for Quantum Federated Learning on Heterogeneous Noisy Hardware}



\author{\IEEEauthorblockN{Walid El Maouaki\IEEEauthorrefmark{1}\IEEEauthorrefmark{2}\orcidlink{0009-0004-2339-5401},
Muhammad Shafique\IEEEauthorrefmark{1}\IEEEauthorrefmark{2}\orcidlink{0000-0002-2607-8135}}

\IEEEauthorblockA{\IEEEauthorrefmark{1} \normalsize eBrain Lab, Division of Engineering, New York University Abu Dhabi, PO Box 129188, Abu Dhabi, UAE\\}
\IEEEauthorblockA{\IEEEauthorrefmark{2} \normalsize Center for Cyber Security, NYUAD Research
Institute, New York University Abu Dhabi, UAE}\\
Emails: walid.el.maouaki@nyu.edu, muhammad.shafique@nyu.edu}
\vspace{-20pt}

\maketitle

\begin{abstract}
Quantum federated learning (QFL) on NISQ hardware is highly sensitive to backend heterogeneity: some clients contribute informative updates, while others contribute noise-dominated drift that uniform averaging cannot distinguish. We propose Q-RAIL (\textit{Quantum Reliability-Aware Federated Inference and Learning}), a circuit- and calibration-aware aggregation method for hardware-heterogeneous QFL. Q-RAIL computes a client-specific effective noise budget from backend calibration metadata together with transpiled circuit statistics. This budget is converted into stabilized aggregation weights using temperature scaling, uniform mixing, and a minimum-weight floor. Q-RAIL was evaluated across multiple experimental settings, including an ablation study, and benchmarked against state-of-the-art methods on three datasets: MNIST, Fashion-MNIST, and OrganAMNIST. On the primary MNIST benchmark under strong hardware skew, Q-RAIL improves final test accuracy from FedAvg's 0.777 to 0.877, a +10.0-point gain corresponding to about 44.8\% relative error reduction, while also exceeding the strongest wpQFL baseline (0.833). At the same time, test loss drops from 0.722 to 0.585, and test AUC rises from 0.920 to 0.973. Under non-IID MNIST, Q-RAIL reaches 0.813 vs 0.722 for FedAvg. It also outperforms FedAvg in 12/12 ansatz/CX-fold stress configurations and remains stronger at 4, 10, and 15 qubit setups. Overall, the results support calibration-driven, circuit-aware aggregation as a practical path toward robust QFL on heterogeneous quantum hardware.
\end{abstract}

\maketitle
\begin{spacing}{0.97}

\vspace{-8pt}
\section{Introduction}
Quantum federated learning (QFL)~\cite{chen2021federated} extends the privacy-preserving promise of federated learning to distributed quantum models, but it also introduces a form of client heterogeneity that is largely absent from classical federated optimization. In classical Federated Learning (FL)~\cite{li2020review}, heterogeneity is usually discussed in terms of non-IID data, variable compute or communication capacity, and optimizer drift across clients. In QFL, however, the same logical model update may be realized on physically different quantum devices whose calibration quality, coherence times, readout fidelity, and qubit connectivity are unequal~\cite{buonaiuto2024effects}. Consequently, two clients that run the same variational quantum model under the same protocol can contribute updates of very different reliability simply because the underlying hardware differs. This setting is particularly relevant in realistic cloud access scenarios, where some participants may only have access to lower-quality backends due to cost, availability, or geographic constraints~\cite{soeparno2021cloud}. FedAvg~\cite{mcmahan2017communication}, FedProx~\cite{li2020federated}, TiFL~\cite{chai2020tifl}, and SCAFFOLD~\cite{karimireddy2020scaffold} all address important forms of heterogeneity in classical FL, but none is designed for the case where the physical faithfulness of a local update depends on backend-specific compilation and execution noise.

\begin{figure}[t]
    \centering
    \includegraphics[width=0.8\linewidth]{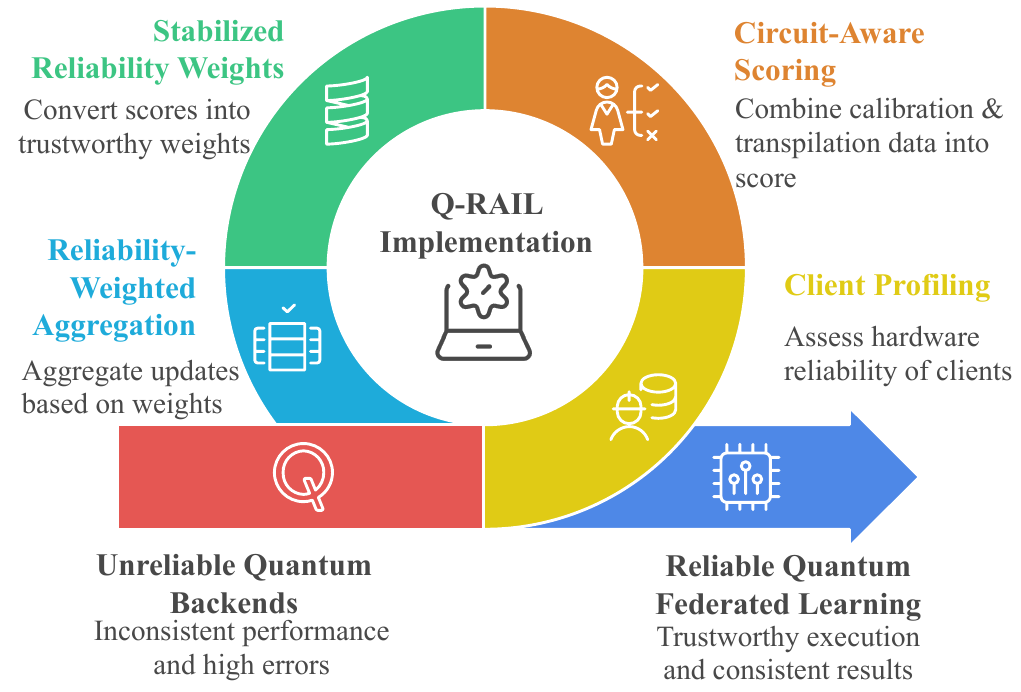}
    \caption{Q-RAIL Implementation Workflow}
    \label{fig:overview}
    \vspace{-15pt}
\end{figure}

This paper studies that regime directly. We define device shift as the discrepancy between a client’s intended logical update and the update that is effectively realized after backend-specific transpilation and noisy execution. Device shift is not determined by nominal error rates alone. In a Noisy Intermediate-Scale Quantum (NISQ)~\cite{preskill2018quantum} setting, the same circuit can compile to different depths, different two-qubit gate counts, and different routing overheads depending on the target backend, because the backend’s target object encodes supported instructions, connectivity constraints, and timing information used by the transpiler~\cite{ibm_transpilation_defaults, ibm_represent_quantum_computers}. Backend properties additionally expose calibration quantities such as one-qubit and two-qubit gate errors, readout error, and coherence indicators derived from $T_1$ and $T_2$~\cite{ibm_view_backend_details}. Therefore, a hardware-aware QFL method should reason jointly about backend quality and the realized circuit that will actually run on that backend.

To address this challenge, we propose Q-RAIL, a noise-aware aggregation method for hardware-heterogeneous QFL. Q-RAIL uses client metadata to estimate an effective noise budget for each client and converts this estimate into a stabilized aggregation factor. Its key insight is that client reliability should not be inferred from backend quality in isolation, but from how faithfully the compiled shared model is expected to execute on that backend. Figure~\ref{fig:overview} presents an workflow of the proposed Q-RAIL.

\noindent\textbf{Contributions:}

\begin{itemize}
    \item Hardware-heterogeneous QFL formulation: We formalize a QFL setting in which client update reliability depends not only on data heterogeneity, but also on backend-specific quantum hardware properties and transpilation outcomes.
    \item Circuit- and calibration-aware reliability scoring: We propose an effective noise budget that combines backend calibration metadata, including gate errors, readout error, and coherence indicators, with transpiled circuit statistics such as depth, one-qubit gates, two-qubit gates, and measurements.
    \item Stabilized reliability-aware aggregation: We introduce Q-RAIL, a server-side aggregation rule that converts effective noise budgets into stable client weights using temperature scaling, uniform mixing, and a minimum-weight floor, giving more influence to reliable clients while preserving participation from noisier devices.
    \item Comprehensive evaluation against QFL baselines: We evaluate Q-RAIL against FedAvg and wpQFL across MNIST, Fashion-MNIST, and OrganAMNIST under IID and non-IID partitions, showing that reliability-aware aggregation improves performance most clearly under strong hardware heterogeneity.
    \item Robustness and scalability analysis: We test Q-RAIL under increasing bad-client ratios, larger client counts, different ansatz topologies, CX-fold noise amplification, and larger qubit regimes, demonstrating that its advantage generally persists beyond the reference 4-qubit setting.
\end{itemize}

\begin{figure*}
    \centering
    \includegraphics[width=\linewidth]{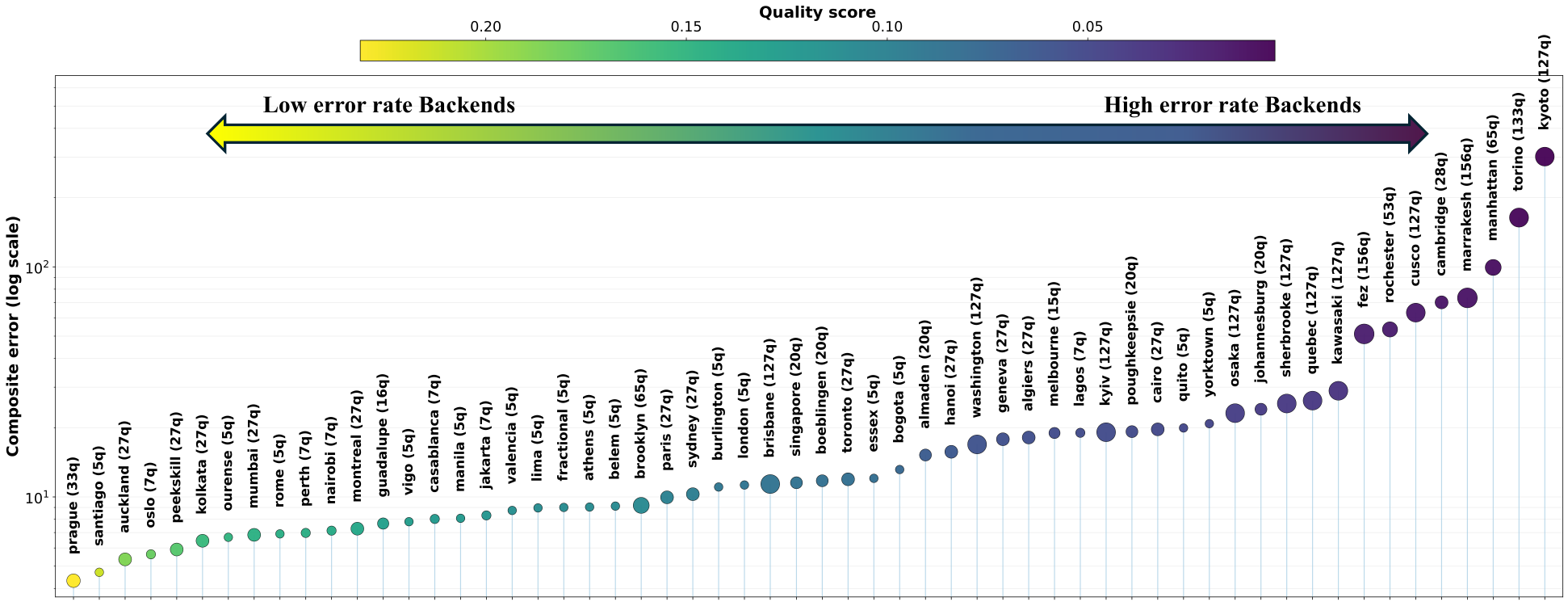}
    \caption{Ranking of candidate backends by composite error (log scale), from low-error to high-error devices, with color indicating the corresponding quality score.}
    \label{fig:backendsScores}
\vspace{-15pt}
\end{figure*}

\section{Background}

FL trains a shared model across distributed clients without centralizing raw data~\cite{mcmahan2017communication}. QFL preserves this decentralized training pattern, but replaces at least part of the local model with a quantum or hybrid quantum-classical model, so the communicated object is typically a set of variational parameters learned from circuit execution rather than a purely classical update~\cite{chehimi2022quantum, ballester2025quantum}. Foundational QFL work and recent reviews therefore position QFL as the intersection of federated learning and quantum machine learning for privacy-preserving distributed model training.


A central difference between classical FL and QFL is that local quantum training is not purely numerical~\cite{liu2025practical}. In Qiskit, the executable circuit is obtained only after transpilation to a backend-specific target, which encodes supported instructions, connectivity, and timing constraints~\cite{ibm_target_v2}. As a result, the same logical ansatz can compile to different depths, different two-qubit gate counts, and different routing overheads across devices~\cite{ibm_intro_transpilation}. This matters because the quantum cost of a client update is backend-dependent even before one considers calibration errors: different coupling maps can induce different routing patterns, and different instruction sets can change the realized gate structure. In short, QFL client heterogeneity is partly a compilation problem before it becomes an optimization problem~\cite{rahman2025toward}.


Quantum noise~\cite{clerk2010introduction} refers to the deviation between intended and realized quantum evolution caused by imperfect control and measurement, including gate errors, readout errors, and decoherence. In NISQ hardware, finite $T_1$ and $T_2$, noisy one- and two-qubit operations, and measurement error accumulate with circuit depth and mapping overhead; Qiskit documentation explicitly notes that shorter circuits generally yield better results because multi-qubit gates are error-prone and qubits decohere over time. 
This issue is especially important for quantum machine learning workload~\cite{biamonte2017quantum}, which is typically implemented as variational or hybrid quantum-classical models. Variational quantum algorithms are designed for near-term quantum devices, but their practical performance remains limited by trainability, accuracy, efficiency, and device noise~\cite{cerezo2021variational}. Prior studies further show that noise can affect variational circuits in a circuit- and noise-dependent manner: some parameterized circuits can partially adapt to noise, but performance may degrade sharply once the noise level exceeds a circuit-dependent threshold~\cite{fontana2021evaluating}. For hybrid quantum neural networks, different noise channels can also have different training effects, with depolarizing noise being particularly harmful in some settings~\cite{kashif2024investigating}. 
For QFL, this means that two clients training the same logical model can still produce updates of unequal reliability if their circuits compile differently or run on backends with different calibration quality; this is precisely the reliability gap that Q-RAIL addresses through circuit- and calibration-aware aggregation.


\section{Related Work}
Classical FL offers several important strategies for heterogeneous learning, but their assumptions do not transfer directly to the quantum setting. FedAvg~\cite{mcmahan2017communication} remains the standard aggregation baseline; FedProx~\cite{li2020federated} stabilizes optimization under systems and statistical heterogeneity; TiFL~\cite{chai2020tifl} addresses client-resource heterogeneity through tiered scheduling; and SCAFFOLD~\cite{karimireddy2020scaffold} corrects client drift under non-IID updates. Personalized and structure-aware methods such as FedMA~\cite{wang2020federated} and FedFA~\cite{zhou2023fedfa} further tackle model or feature heterogeneity. However, these approaches still assume that a client’s local computation is a faithful realization of the intended numerical update. They do not account for target-specific transpilation, topology-induced SWAP overhead, or calibration-conditioned gate fidelity. In QFL, those effects alter the reliability of the update itself, which is why classical heterogeneity remedies cannot simply be copied into the quantum layer without additional hardware-aware modeling.


Within QFL, early work such as quantum federated learning with quantum data~\cite{chehimi2022quantum} established the feasibility of federated quantum training and motivated the broader study of distributed quantum models. More recent personalized QFL work, including the weighted personalized QFL (wpQFL)~\cite{gurung2025performance} baseline used in this paper, addresses non-IID data and local model drift using weighted or personalized qFedAvg-style strategies. That baseline is a strong comparison point precisely because it improves on standard aggregation under data heterogeneity; in our study, it is reproduced using the original protocol and the settings reported as strongest in that work. Nevertheless, its weighting logic is not based on backend quality or transpiled-circuit reliability, which leaves hardware heterogeneity only indirectly addressed, if at all.

Very recent QFL literature has started to confront hardware heterogeneity more directly. SpoQFL~\cite{rahman2025sporadic} introduces sporadic learning to mitigate quantum-noise heterogeneity across devices, and SPQFL~\cite{rahman2026tackling} extends this direction by combining sporadic learning for noise heterogeneity with personalization for non-IID data. 
These methods, however, differ from Q-RAIL in both objective and mechanism: they respond to noisy heterogeneity by modifying local training or participation to clean, suppress, or skip unreliable client updates, whereas Q-RAIL keeps all participating updates and performs server-side reliability-weighted aggregation using backend metadata and the actual transpiled circuit. This preserves data diversity that skipped updates may lose and makes the method directly applicable to current NISQ hardware.

Finally, Q-RAIL is complementary to adversarially robust QFL. QFAL~\cite{maouaki2025qfal} and RobQFL~\cite{el2025robqfl} study how QFL behaves under adversarial perturbations and adversarial training regimes. Those works are important because they show that QFL is sensitive not only to data skew but also to malicious or adversarial conditions~\cite{el2025designing}. However, their threat model is adversarial manipulation, whereas the present paper studies benign but systematic client-quality inequality induced by heterogeneous hardware access. In that sense, Q-RAIL is not a replacement for QFAL or RobQFL; it addresses an orthogonal reliability problem that could, in principle, later be combined with adversarially robust QFL methods~\cite{el2024advqunn, el2025designing}.

\section{Methodology}

\subsection{Federated setting and notation}

We consider a federation of $K$ quantum clients. Client $k \in \{1, \ldots, K\}$ holds a private dataset
\[
\mathcal{D}_k = \left\{ \left(x_i^{(k)}, y_i^{(k)}\right) \right\}_{i=1}^{n_k},
\]
and is assigned a backend $b_k$ from a candidate pool $\mathcal{B}$. All clients share the same logical variational quantum classifier (VQC) $f(x;\theta)$ with global parameter vector $\theta^t \in \mathbb{R}^P$ at communication round $t$, but they execute backend-specific transpiled circuits $T_{b_k}(f)$. The server receives local parameter deltas
\[
\Delta_k^t = \theta_k^{t,\mathrm{loc}} - \theta^t,
\]
and computes the next global model through a weighted aggregation rule. Q-RAIL differs from uniform FedAvg by making these weights depend on a client-specific estimate of update reliability derived from the client's backend properties and transpiled circuit statistics. The server is assumed to receive backend properties and transpiled gate-count metadata from each client; Q-RAIL does not require access to raw data or low-level pulse traces.

\subsection{Local quantum model and preprocessing}

Each input sample is first reduced to $Q$ real-valued features and then scaled to $[0, \pi]$. We denote the encoded feature vector by $x = (x_1, \ldots, x_Q)$. The feature map is
\[
\Phi(x) = \prod_{j=1}^{Q} R_y(x_j) R_z(x_j),
\]
which implements a one-feature-per-qubit angle encoding. 
The trainable circuit then applies $L$ repeated layers of single-qubit rotations followed by an entangling block:
\[
U(x;\theta) = \left( \prod_{\ell=1}^{L} E_{\ell}^{(\tau,f)} \prod_{j=1}^{Q} R_y(\theta_{\ell,2j-1}) R_z(\theta_{\ell,2j}) \right) \Phi(x),
\]
where $\tau \in \{\mathrm{linear}, \mathrm{ring}, \mathrm{full}\}$ denotes the entanglement topology and $f$ is the CX-fold parameter. The total number of trainable parameters is therefore
$
P = 2QL.
$

For the primary benchmark $Q = 4$ and $L = 4$, so $P = 32$. This reference model is intentionally small enough to execute across many candidate backends while still containing repeated entangling structure that exposes the effect of hardware noise. It is a controlled benchmark, not a claim of universal optimality. A second regime with $Q = 8$ and $L = 16$ is included to match the scale used by the wpQFL baseline. Later ablations vary $Q$, $\tau$, and $f$ to test generality.

In the implementation, the entangling block supports linear, ring, and dense all-to-all (full) connectivity, and CX folding repeats each entangling gate an odd number of times with barriers inserted to prevent cancellation during transpiler optimization. This is important for the stress-test experiments, because it creates a controlled way to amplify the exposure of the circuit to two-qubit noise and routing overhead. For the three-class tasks used here, the model measures two qubits and maps the bitstrings 00, 01, and 10 to the three labels. Local training uses SPSA, which is appropriate for shot-based quantum optimization because it estimates gradients from two function evaluations rather than requiring analytic differentiation of noisy circuits. The SPSA update at local step $s$ can be written as $\theta_{k,s+1} = \theta_{k,s} - a_s \hat{g}_{k,s}$,
\[
\hat{g}_{k,s} =
\frac{\mathcal{L}(\theta_{k,s} + c_s \Delta_{k,s}) - \mathcal{L}(\theta_{k,s} - c_s \Delta_{k,s})}{2c_s}\,\Delta_{k,s},
\]
with $\Delta_{k,s} \in \{-1, +1\}^{P}$, $a_s = a_0/(s+1)^{\alpha}$, and $c_s = c_0/(s+1)^{\gamma}$. In the current implementation, the reference values are $a_0 = 0.2$, $c_0 = 0.1$, $\alpha = 0.602$, and $\gamma = 0.101$, with 5 local steps, batch size 16, 128 training shots, and 256 evaluation shots.

\begin{figure*}[t]
    \centering
    \includegraphics[width=\linewidth]{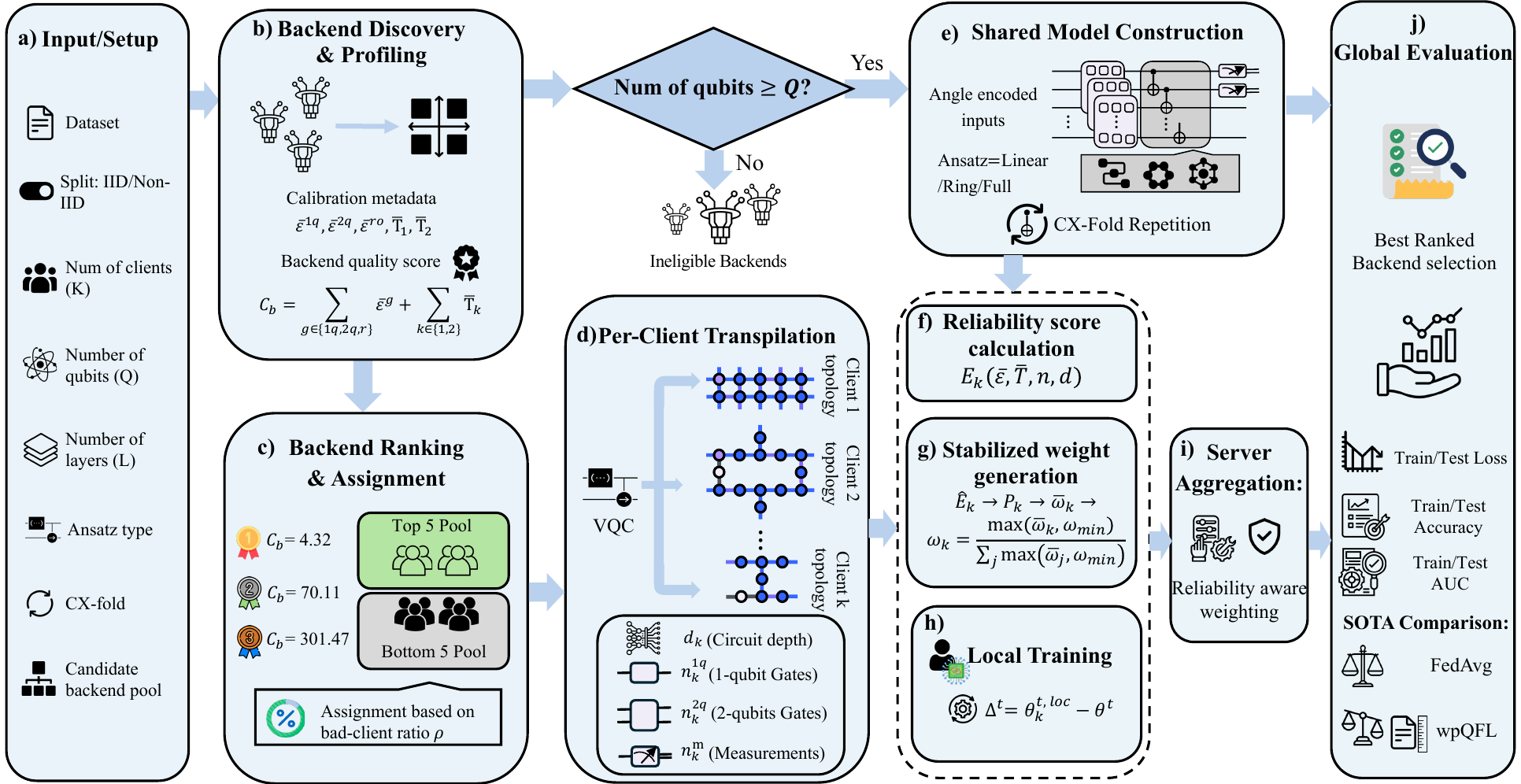}
    \caption{Overview of the Q-RAIL methodology: a) Input. b) Backend discovery and profiling. c) Ranking and client assignment. Here, the top-5 and bottom-5 pools are shown as an example, but the pool size is adjustable based on the full candidate backend set. d) Per-client transpilation. e) Shared QML model construction. f) Reliability-score computation. g) Stabilized reliability-aware aggregation. h) Local training. i) Server aggregation. j) Global evaluation and SOTA comparison.}
    \label{fig:Methodology}
    \vspace{-20pt}
\end{figure*}

\subsection{Backend characterization and client assignment}

Let
\[
\mathcal{B}_Q = \{b \in \mathcal{B} : \mathrm{num\_qubits}(b) \ge Q\}
\]
be the set of fake backends that can host a $Q$-qubit circuit. For each $b \in \mathcal{B}_Q$, we extract the average one-qubit gate error $\bar{\epsilon}_b^{1q}$, average two-qubit gate error $\bar{\epsilon}_b^{2q}$, average readout error $\bar{\epsilon}_b^{ro}$, and average coherence indicators $\bar{T}_{1,b}$ and $\bar{T}_{2,b}$. These quantities are available through backend properties, while transpilation itself uses the target object and backend constraints. We then define a backend-only composite score
\[
C_b = \lambda_{2q}\,\widetilde{\bar{\epsilon}_b^{2q}} + \lambda_{1q}\,\widetilde{\bar{\epsilon}_b^{1q}} + \lambda_{ro}\,\widetilde{\bar{\epsilon}_b^{ro}} + \lambda_{T1}\,\widetilde{\bar{T}_{1,b}^{-1}} + \lambda_{T2}\,\widetilde{\bar{T}_{2,b}^{-1}},
\]
where $\tilde{\cdot}$ denotes median normalization over $\mathcal{B}_Q$. Lower $C_b$ indicates better backend quality. In the default implementation, $(\lambda_{1q}, \lambda_{2q}, \lambda_{ro}, \lambda_{T1}, \lambda_{T2}) = (1, 5, 2, 1, 1)$, giving the strongest prior weight to two-qubit error.

The candidate backends are then sorted by $C_b$, and the framework forms two assignment pools by splitting them into the better-performing half and the worse-performing half. In Figure~\ref{fig:Methodology}, the partition size is set to 5 for illustrative purposes; however, this value is configurable and can be adjusted based on the total number of available backends to ensure an appropriate division into the two pools. Figure~\ref{fig:backendsScores} list the actual fake backends available in Qiskit~\cite{ibm_fake_provider_backend_v2}. Client assignment is randomized within this filtered regime: for a chosen bad-client ratio $\rho$, approximately $\rho K$ clients are sampled from the low-quality pool and the remaining clients from the high-quality pool, after which the assignments are shuffled. This design makes the experimental notion of ``good'' and ``bad'' hardware precise, while still varying backend identity stochastically across runs. Table~\ref{tab:BackendsDistribution} presents an example of client assignments for 80\% of the bad clients setting.

\input{Tables/transpilation}

\subsection{Circuit-aware effective noise budget}

Backend ranking alone is insufficient for aggregation, because it does not account for how a particular model compiles on a particular backend. For each client $k$, we therefore transpile the shared circuit to $b_k$ and extract
\[
d_k,\quad n_k^{1q},\quad n_k^{2q},\quad n_k^{m},
\]
respectively the transpiled depth, the number of one-qubit gates, the number of two-qubit gates, and the number of measurement operations.. Let the corresponding average backend properties be $\bar{\epsilon}_k^{1q}$, $\bar{\epsilon}_k^{2q}$, $\bar{\epsilon}_k^{ro}$, $\bar{T}_{1,k}$, and $\bar{T}_{2,k}$. Q-RAIL defines the raw execution-risk components
\[
r_k^{2q} = n_k^{2q}\bar{\epsilon}_k^{2q}, \qquad
r_k^{1q} = n_k^{1q}\bar{\epsilon}_k^{1q}, \qquad
r_k^{ro} = n_k^{m}\bar{\epsilon}_k^{ro},
\]
\[
r_k^{T1} = \frac{d_k}{\bar{T}_{1,k} + \epsilon}, \qquad
r_k^{T2} = \frac{d_k}{\bar{T}_{2,k} + \epsilon}.
\]

Each component is median-normalized across participating clients to prevent scale domination:
\[
\tilde{r}_k^s = \frac{r_k^s}{\mathrm{median}_j(r_j^s) + \epsilon}, \qquad
s \in \{1q, 2q, ro, T1, T2\}.
\]

The final effective noise budget is then
\[
E_k = \lambda_{2q}\tilde{r}_k^{2q} + \lambda_{1q}\tilde{r}_k^{1q} + \lambda_{ro}\tilde{r}_k^{ro} + \lambda_{T1}\tilde{r}_k^{T1} + \lambda_{T2}\tilde{r}_k^{T2}.
\]

This quantity is a first-order reliability score that combines calibration quality with the realized circuit footprint. Q-RAIL estimates how trustworthy a client update is likely to be after compilation on that backend.

\subsection{Q-RAIL aggregation rule}

Given the set of effective noise budgets $\{E_k\}_{k=1}^{K}$, Q-RAIL maps them to aggregation weights through a three-stage stabilization rule. First, budgets are normalized to $[0,1]$:
\[
\hat{E}_k = \frac{E_k - \min_j E_j}{\max_j E_j - \min_j E_j + \epsilon},
\]
so that lower values correspond to more reliable clients. Second, a temperature-controlled softmax is applied:
\[
p_k = \frac{\exp(-\tau \hat{E}_k)}{\sum_{j=1}^{K} \exp(-\tau \hat{E}_j)}.
\]

Third, this distribution is regularized by mixing with uniform weights and applying a floor:
\[
\bar{w}_k = (1-\beta)p_k + \frac{\beta}{K}, \qquad
w_k = \frac{\max(\bar{w}_k, w_{\min})}{\sum_{j=1}^{K} \max(\bar{w}_j, w_{\min})}.
\]

In the default implementation, $\tau = 5.0$, $\beta = 0.2$, and $w_{\min} = 0.05$. The server update is then
\[
\theta^{t+1} = \theta^t + \sum_{k=1}^{K} w_k \Delta_k^t.
\]

This stabilization is important. A pure reliability softmax assigns higher normalized weights to cleaner clients and lower weights to noisier ones, with temperature controlling the sharpness of this preference.
The uniform mixture guarantees that all clients retain some participation, while the minimum-weight floor prevents any client from being driven to a near-zero influence. Thus, the method maintains diversity and robustness in the federation while giving cleaner, more reliable clients greater influence on the global update. Relative to FedAvg, the only structural change is the introduction of reliability-aware weights. Relative to wpQFL, Q-RAIL places the weighting signal in the hardware-and-circuit domain rather than in model-parameter distance alone. 
Figure~\ref{fig:Methodology} provides a comprehensive overview of the proposed Q-RAIL framework.

\subsection{Workflow}

Algorithm~\ref{alg:Alg1} summarizes the end-to-end Q-RAIL workflow used in the implementation. Lines 1--5 discover eligible fake backends and compute the backend-only composite score $C_b$; lines 6--10 assign client devices within the selected qubit regime, instantiate the shared VQC, and compute the circuit-aware effective noise budget $E_k$ for each client; and lines 11--24 perform round-wise local SPSA training, compute the client deltas $\Delta_k^t$, derive the Q-RAIL aggregation weights $w_k^t$, update the global parameters, and evaluate the resulting global model on the best-ranked backend $b^*$. This matches the current code path, which ranks fake BackendV2 devices, forms top-5 and bottom-5 pools, samples client assignments according to the bad-client ratio, computes $E_k$ from transpiled metrics and backend properties, and records global, local, and weight-level outputs after each round.

\input{Algorithms/Algorithms1}

Within this shared framework, only the aggregation step changes across compared methods. FedAvg sets $w_k^t = 1/K$, Q-RAIL uses the reliability-aware rule in lines 18--22, and wpQFL replaces the weighting step with its original personalized update rule while keeping the remaining pipeline fixed. This design ensures that differences in performance are attributable to aggregation rather than to changes in preprocessing, local optimization, backend assignment, or evaluation protocol.

\section{Experimental Setup}

The primary benchmark, \textbf{Configuration A}, uses a 4-qubit, 4-layer angle-encoded VQC over 10 clients and serves as the reference regime for isolating hardware effects. This configuration is not claimed to be universally representative; rather, it is the controlled benchmark used throughout the main study. A second \textbf{Configuration B} uses 8 qubits and 16 layers to mirror the scale used in the weighted personalized QFL baseline for direct comparison. Local training uses SPSA with 5 local steps per round, batch size 16, 128 training shots, and 256 evaluation shots. We compare Q-RAIL against \textbf{FedAvg} and an exact implementation of \textbf{wpQFL}, using the Euclidean setting and the weighted 90g10l setting, where the personalized update is formed with 90\% global and 10\% local contribution, as reported among the strongest configurations in the original study. Reported metrics include train/test loss, train/test accuracy, train/test AUC, and per-device average metrics.

The Q-RAIL implementation uses an angle-encoded variational classifier. Inputs are first reduced by PCA to $Q$ features and then scaled to $[0,\pi]$, after which each feature is mapped to one qubit through $\mathrm{RY}$ and $\mathrm{RZ}$ rotations. This choice is methodologically important. Angle encoding provides a direct one-feature-per-qubit mapping and keeps the feature-map depth fixed; consequently, most of the depth variation can be attributed to the trainable ansatz and to backend-dependent transpilation rather than to a growing data-encoding block. This makes hardware effects easier to interpret and control. The default 4-qubit setting is therefore not arbitrary: it matches a four-dimensional reduced feature representation and yields $2QL = 32$ trainable parameters in the 4-layer reference model, which is expressive enough for a nontrivial classifier while remaining executable on a wide set of candidate backends. Broader representativeness is then examined through the 8-qubit/16-layer comparison setting and the later qubit- and ansatz-ablation experiments.

We use Qiskit fake backends from FakeProviderForBackendV2~\cite{ibm_fake_provider_backend_v2} as fixed hardware proxies, since they reproduce the backend features needed for transpilation and noisy execution. They come from historical snapshots of real devices, which suits Q-RAIL’s need for controlled heterogeneous device states. Backend properties expose calibration-derived quantities such as $T_1$ and $T_2$, readout error, and instruction-level errors, while the target stores connectivity and per-instruction properties used during transpilation. IBM documentation further notes that family-specific two-qubit primitives can differ, for example with CZ used on some families and ECR on others, which reinforces the point that “hardware heterogeneity” in QFL is not just a matter of one scalar noise level. The actual backends used in this work, their qubit counts, and their computed quality scores are reported in Figure \ref{fig:backendsScores}. Client backends are then sampled randomly, within the relevant qubit regime, from the selected high-quality and low-quality pools. All experiments were conducted using Qiskit 2.3.0 and Qiskit IBM Runtime 0.43.1.

Table~\ref{tab:backend-transpilation} shows the impact of backend-specific compilation on the same 5-qubit, 12-layer ring-entangled VQC. Across backends and optimization levels, transpiled depth ranges from 209 to 675, two-qubit gates from 184 to 252, and routing adds 45–66 SWAPs. At optimization level~1, depth increases by \(2.4\times\)–\(6.7\times\) and the two-qubit gate count by \(3.25\times\)–\(4.20\times\). Thus, backend topology, native gates, and routing constraints substantially alter the executed circuit and its client-specific noise exposure in Q-RAIL.

\input{Tables/ExpSetup}


We performed multiple experiments to evaluate the proposed method. The first one is the \textbf{main comparison}, where Q-RAIL, FedAvg, and wpQFL are compared on the primary benchmark (Configuration A) using MNIST~\cite{lecun1998gradient}, Fashion-MNIST~\cite{xiao2017fashionmnist}, and OrganAMNIST~\cite{yang2023medmnistv2, yang2021medmnistv1} under both IID and non-IID partitions, generated using a Dirichlet distribution to induce label skew~\cite{ferguson1973dirichlet}. The experiments use 8,000/1,600 train/test samples for MNIST and Fashion-MNIST, and 4,700/1,000 train/test samples for OrganAMNIST.
A second comparison within this main experiment uses Configuration B on MNIST under IID and non-IID settings to provide a like-for-like comparison at the scale used in wpQFL. 

The second experiment is a \textbf{bad-client sweep}: using Config A, Q-RAIL is compared with FedAvg and wpQFL while varying the fraction of low-quality clients, $\rho \in {0.2, 0.5, 0.8, 1.0}$, in order to measure how well the aggregation rule tolerates increasingly adverse hardware composition. The third experiment is a \textbf{client-count sweep}, where, again using Config A, the number of clients is varied over $K \in {10, 20, 50}$ to test whether the relative advantage of Q-RAIL persists as the federation becomes larger and more heterogeneous, Q-RAIL is compared with FedAvg and wpQFL on IID MNIST data. The fourth experiment is an \textbf{ansatz and CX-fold stress test}: using Config A, the entanglement topology is varied across Linear, Ring, and Full, while the CX-fold parameter is varied over ${1, 3, 5, 9}$ folds. 
Q-RAIL is compared with FedAvg on IID MNIST data. Finally, the fifth experiment is a \textbf{qubit-scale study}, in which Q-RAIL and FedAvg are compared at $Q \in {4, 10, 15}$ qubits to study how the proposed aggregation rule behaves as circuit size increases, Q-RAIL is compared with FedAvg and wpQFL on IID MNIST data. Unless stated otherwise, the proportion of bad clients in all experiments is set at 80\%.

The federation is performed using 15 rounds, and the experiments are repeated over five random seeds, and we report the mean performance with standard deviations, or mean performance with 95\% confidence intervals. Table~\ref{tab:expsetup} presents an overview of the experimental setup used in this study.

\input{Tables/table0}

\input{Tables/table1}
\input{Tables/table2}

\begin{figure*}
    \centering
    \includegraphics[width=1\linewidth]{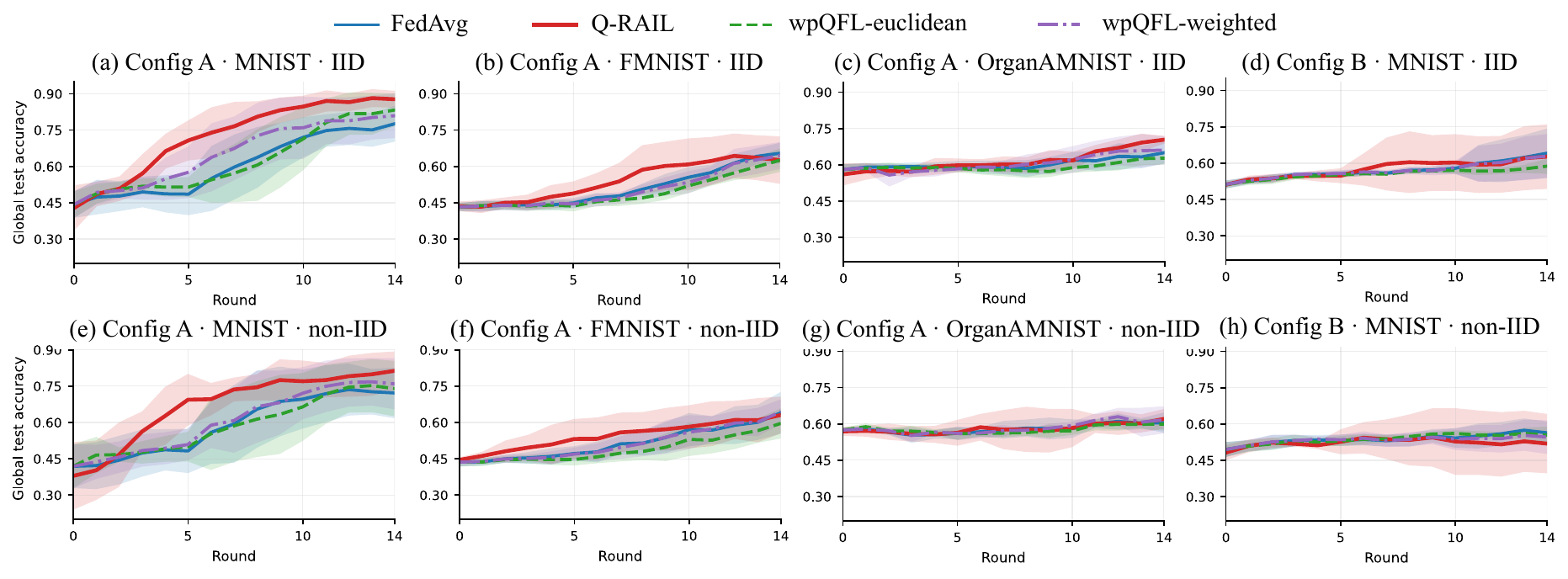}
    \caption{Main comparison across Configurations A and B. Global test accuracy over communication rounds for FedAvg, Q-RAIL, and wpQFL under IID (top row) and non-IID (bottom row) data partitions on MNIST, Fashion-MNIST, and OrganAMNIST. The legend is shared.}
    \label{fig:maincompa}
    \vspace{-10pt}
\end{figure*}

\begin{figure}
    \centering
    \includegraphics[width=1\linewidth]{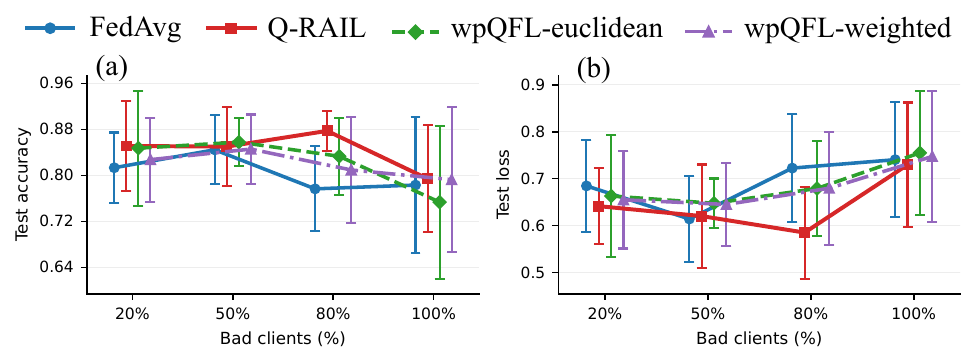}
    \caption{Performance under increasing bad-client ratios. Final-round test accuracy and test loss as the proportion of low-quality clients increases from 20\% to 100\%. Error bars denote variability across seeds. The legend is shared.}
    \label{fig:badsweep}
    \vspace{-10pt}
\end{figure}

\begin{figure}
    \centering
    \includegraphics[width=1\linewidth]{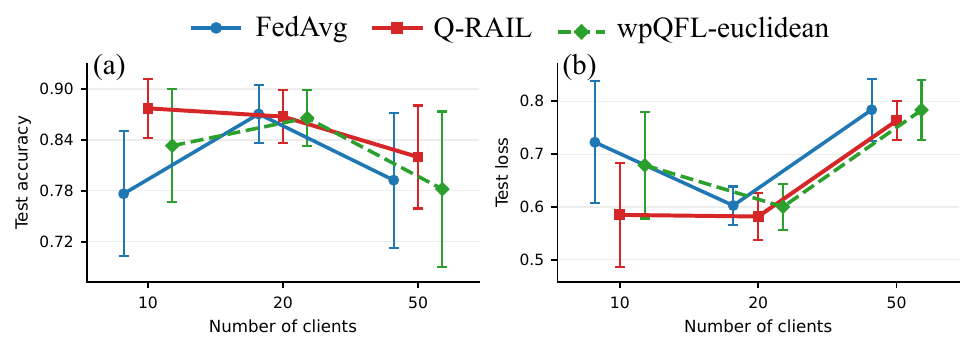}
    \caption{Client-count sweep on IID MNIST. Final-round test accuracy and test loss for FedAvg, Q-RAIL, and wpQFL as the number of clients increases from 10 to 50. The legend is shared.}
    \label{fig:ClientsSweep}
    \vspace{-10pt}
\end{figure}

\begin{figure}
    \centering
    \includegraphics[width=1\linewidth]{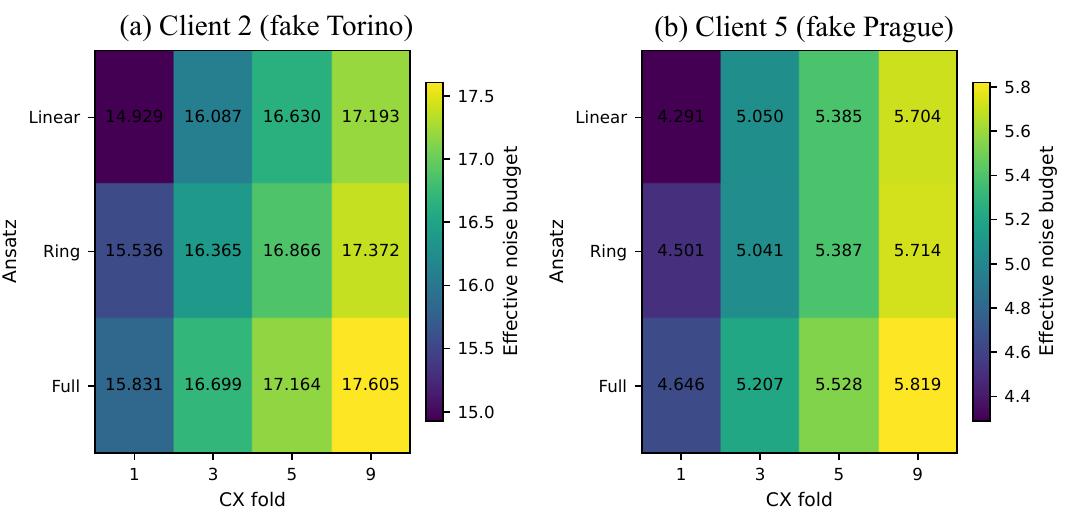}
    \caption{Representative Q-RAIL effective noise budgets. Example effective-noise-budget heatmaps for representative clients across ansatz topologies and CX-fold values. In both views, the estimated budget increases with stronger entanglement and repeated CX insertion.}
    \label{fig:AnsatzCXFoldNoise}
    \vspace{-10pt}
\end{figure}

\begin{figure}
    \centering
    \includegraphics[width=1\linewidth]{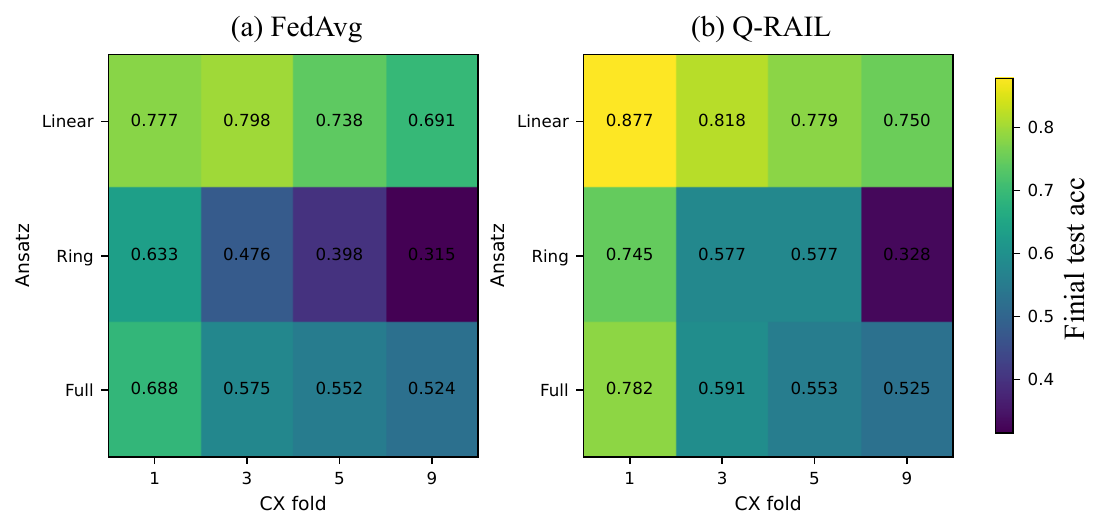}
    \caption{Ansatz and CX-fold stress test. Final test accuracy of FedAvg and Q-RAIL across linear, ring, and full entanglement patterns under increasing CX-fold values. Higher CX-fold amplifies circuit noise by repeating entangling operations.}
    \label{fig:AnsatzCXFoldAccuracy}
    \vspace{-10pt}
\end{figure}

\section{Results}
\subsection{Main comparison}

Figure~\ref{fig:maincompa} presents the main comparison results. Q-RAIL achieves the best performance on the primary benchmark, Configuration A, with its most pronounced gains observed on MNIST. In the MNIST-IID setting, Q-RAIL attains a test accuracy of $0.877$, outperforming FedAvg, which reaches $0.777$, wpQFL (Euclidean) with $0.833$, and wpQFL (weighted) with $0.809$; it also gives the lowest test loss (0.585) and highest test AUC (0.973). The same pattern holds for MNIST-non-IID, where Q-RAIL attains 0.813 test accuracy vs 0.722 (FedAvg), 0.740 (wpQFL-euclidean), and 0.760 (wpQFL-weighted), Table~\ref{tab:mainComparison} reports the results for the main comparison across all methods and settings. Q-RAIL shows faster improvement and reaches a higher final accuracy. OrganAMNIST shows smaller but still consistent gains, with Q-RAIL giving the highest test accuracy in both IID (0.704) and non-IID (0.621) settings; see Table~\ref{tab:mainComparison}. The learning curves also show faster separation of Q-RAIL from the baselines after only a few rounds, see Figure~\ref{fig:maincompa}a), c), e), and g).

On the FMNIST data, the IID results are somewhat mixed. Although FedAvg achieves a slightly higher test accuracy (0.654) than Q-RAIL (0.626), Q-RAIL obtains the highest test AUC (0.832) among all compared techniques. Moreover, Q-RAIL still surpasses wpQFL in test accuracy, and in the non-IID setting it emerges as the best-performing method overall, this is illustrated in Table~\ref{tab:mainComparison}, where the best-performing results are indicated in boldface. In Figure~\ref{fig:maincompa}b) and f), Q-RAIL maintains a clear performance advantage after the early rounds.
In the larger Configuration B, the advantage of Q-RAIL becomes less pronounced: under IID, the methods remain close throughout training, while under non-IID all methods degrade substantially, and Q-RAIL no longer leads in test accuracy, even though it remains competitive in loss and AUC, see Figure~\ref{fig:maincompa}d) and h), and Table~\ref{tab:mainComparison}. Overall, the main comparison indicates that Q-RAIL is most beneficial when hardware heterogeneity is strong but the task remains sufficiently learnable.

\subsection{Bad-client sweep}

Figure~\ref{fig:badsweep} summarizes the robustness of each method under increasing bad-client ratios, reporting final-round test accuracy and loss with variability across seeds. Q-RAIL becomes more beneficial as the proportion of low-quality devices increases. At 20\% bad clients, Q-RAIL achieves the best test accuracy (0.851) and lowest test loss (0.641), improving over FedAvg by $+3.8$ points in accuracy. At 50\% bad clients, wpQFL (euclidean) is slightly best in test accuracy and AUC (0.858 / 0.977), while Q-RAIL remains close in test accuracy (0.850). The strongest result appears at 80\% bad clients, where Q-RAIL clearly dominates: test accuracy rises to 0.877, versus 0.777 for FedAvg, 0.833 for wpQFL (euclidean), and 0.809 for wpQFL (weighted), while test loss drops to 0.585. Even at 100\% bad clients, Q-RAIL remains best or tied-best in test accuracy (0.794) and is best in test loss (0.730) and test AUC (0.925). Tables~\ref{tab:BadClientsSweep} reports the corresponding training and testing performance metrics, where the best-performing results are shown in bold.

The bad-client local metrics show the same trend. At 80\% bad clients, Q-RAIL increases bad-client local accuracy to 0.772, compared with 0.684 for FedAvg, 0.731 for wpQFL (euclidean), and 0.727 for wpQFL (weighted), see Tables~\ref{tab:BadClientsSweep}. Thus, the gain is not only at the server level; Q-RAIL also improves the average quality of updates coming from the weakest hardware group.

\subsection{Client-count sweep}

Figure~\ref{fig:ClientsSweep} depicts the test accuracy and test loss achieved by the respective methods as a function of the number of clients. When the number of clients increases from 10 to 50, Q-RAIL remains competitive and usually achieves the best performance. At 10 clients, it gives the highest test accuracy and lowest test loss. At 20 clients, all methods are close, with only a small gap between Q-RAIL and wpQFL. At 50 clients, Q-RAIL again achieves the best accuracy and a lower loss than FedAvg, while remaining slightly better than wpQFL in test accuracy. Hence, the benefit of Q-RAIL does not disappear as the federation grows, although the margin becomes smaller as client diversity increases.

\subsection{Ansatz and CX-fold stress test}

Figure~\ref{fig:AnsatzCXFoldNoise} illustrates heatmaps of how the Ansatz entanglement topology and the number of fold layers influence the effective noise of the transpiled circuit on two representative client backends for the same VQC. Client 2 is associated with the noisiest backend in the pool (highest composite error), while Client 5 is associated with the least noisy backend (lowest composite error).
Figure~\ref{fig:AnsatzCXFoldAccuracy} shows the heatmaps of the final test accuracy achieved by FedAvg and Q-RAIL as a function of the Ansatz entanglement topology and the number of fold layers.

As expected in Figure~\ref{fig:AnsatzCXFoldAccuracy}a) and b), increasing CX-fold reduces performance for both methods, and denser entanglement generally leads to harder settings. Still, Q-RAIL outperforms FedAvg in nearly all cases. For the linear ansatz, Q-RAIL improves final test accuracy at every fold: 0.877 vs 0.777 (fold 1), 0.818 vs 0.798 (fold 3), 0.779 vs 0.738 (fold 5), and 0.750 vs 0.691 (fold 9). For the ring ansatz, the gains are even larger at moderate stress: 0.745 vs 0.633 (fold 1), 0.577 vs 0.476 (fold 3), and 0.577 vs 0.398 (fold 5), before narrowing at fold 9. For full entanglement, Q-RAIL is clearly better at fold 1 and approximately tied at higher folds. These results show that Q-RAIL is especially effective when circuit noise is amplified but not yet fully overwhelming.

This interpretation is supported by the representative effective-noise-budget heatmaps, Figure~\ref{fig:AnsatzCXFoldNoise}a) and b). For the representative clients shown, the Q-RAIL effective noise budget increases monotonically with CX-fold, and it also tends to increase as the ansatz becomes more entangling. This mirrors the observed drop in accuracy and confirms that the weighting signal used by Q-RAIL tracks circuit difficulty in the intended direction.

For the more severe 5-fold and 9-fold noise settings, we changed the noise-aware parameters from $\tau = 5.0$, $\beta = 0.2$, and $w_{\min} = 0.05$ to $\tau = 10$, $\beta = 0.05$, and $w_{\min} = 0.01$. This makes the weighting more discriminative by strengthening the softmax preference for cleaner clients and reducing the stabilizing pull toward uniform participation. The change is necessary because, under higher effective noise, the default settings are overly smoothed and allow poor backends to perturb overall performance specifically in the 80\% of bad client setting.

\subsection{Qubit-scale study}

\begin{figure}
    \centering
    \includegraphics[width=0.6\linewidth]{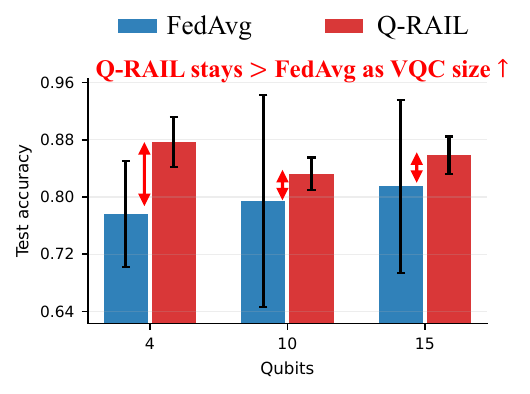}
    \caption{Qubit-scale study. Final test accuracy of FedAvg and Q-RAIL across VQC sizes of 4, 10, and 15 qubits, showing that the advantage of Q-RAIL persists beyond the reference 4-qubit setting. Error bars denote variability across seeds.}
    \label{fig:QubitsScale}
\vspace{-10pt}
\end{figure}

Figure~\ref{fig:QubitsScale} presents a bar chart that compares the mean test accuracy of Q-RAIL and FedAvg as a function of the number of qubits. Q-RAIL outperforms FedAvg at all tested qubit counts. The gap is largest at 4 qubits (about 0.88 vs 0.78 test accuracy), and it remains positive at 10 qubits (about 0.84 vs 0.79) and 15 qubits (about 0.86 vs 0.81). While the absolute margin changes with scale, the result suggests that the Q-RAIL weighting mechanism remains useful beyond the reference 4-qubit setting.
\vspace{-8pt}

\section{Discussion}
The results support the central claim of the paper: in hardware-heterogeneous QFL, aggregation should depend on update reliability, not only on client participation. Q-RAIL is strongest precisely when hardware skew is most harmful, namely in the 80\%-100\% bad-client regimes, and in the CX-fold stress tests. In these settings, weighting clients by backend properties together with transpiled circuit cost yields clear improvements in accuracy and reductions in loss relative to uniform aggregation.

The bad-client sweep is especially important because it isolates the effect of hardware heterogeneity. When only 20\% of clients are poor, Q-RAIL is already best, but the largest gains appear at 80\% bad clients, where uniform aggregation becomes most vulnerable to low-quality updates. This is exactly the regime Q-RAIL is designed for: it reduces the influence of physically unreliable updates rather than assuming that all client updates should contribute equally.

The comparison with wpQFL is also informative. wpQFL remains competitive when the heterogeneity is moderate, especially at 50\% bad clients and around 20 clients, but Q-RAIL is more robust once the heterogeneity is explicitly hardware-driven and severe. This difference is consistent with the design of the methods: wpQFL reweights or personalizes using parameter-space relations, whereas Q-RAIL uses backend properties and transpiled gate statistics that directly reflect quantum execution risk.

At the same time, the gains are not universal. On FMNIST, the improvements are small and sometimes absent, and under Configuration B non-IID Q-RAIL does not achieve the highest test accuracy. This suggests two practical limits. First, when the learning problem itself is harder, the benefit of reliability-aware aggregation can be masked by optimization difficulty. Second, when circuits become too deep or all clients are similarly noisy, reweighting alone cannot fully recover the information lost to execution noise. The stress-test results support this reading: Q-RAIL remains helpful under increasing CX-fold, but its margin narrows in the most extreme high-noise settings.

Finally, the ansatz/CX-fold and qubit-scale experiments strengthen the methodological rationale of Q-RAIL. The effective-noise-budget heatmaps show a monotonic rise in estimated execution risk as entanglement density and CX repetition increase, and the accuracy heatmaps show corresponding degradation in final performance. Together, these results indicate that the Q-RAIL weighting signal is aligned with circuit difficulty rather than being an arbitrary heuristic. The qubit-scale study further suggests that this benefit is not restricted to the smallest circuit, even if the margin varies with scale. In short, Q-RAIL improves robustness to backend heterogeneity, but it does not eliminate noise; its main strength lies in the realistic middle regime where client updates differ meaningfully in trustworthiness.

\vspace{-0.2cm}
\section{Conclusion}
\vspace{-0.1cm}
This paper introduced Q-RAIL for hardware-heterogeneous quantum federated learning. In QFL, client updates should not be trusted equally when they are produced on backends with different calibration quality and compiled circuit costs. Q-RAIL addresses this by combining backend properties with transpiled circuit statistics into an effective noise budget and then using a stabilized reliability-aware weighting rule for server aggregation. Across the main experiments, the method delivers its strongest gains on the MNIST benchmark, where test accuracy improves from 0.777 to 0.877 under IID and from 0.722 to 0.813 under non-IID relative to FedAvg, while also outperforming the strongest wpQFL setting in both cases. It is especially effective when hardware imbalance is severe: at 80\% bad clients, Q-RAIL yields +10.0 points in test accuracy, reduces test loss from 0.722 to 0.585, improves AUC from 0.920 to 0.973, and increases bad-client local accuracy from 0.684 to 0.772. Under ansatz/CX-fold stress testing, it exceeds FedAvg in all 12 settings, and it remains above FedAvg as the qubit count increases from 4 to 15.

These findings are important because they show that device heterogeneity is not a secondary implementation detail in QFL; it is part of the learning problem itself. The results suggest that aggregation based only on uniform averaging or parameter-space relations, is insufficient when client updates differ in physical reliability. At the same time, the experiments also show the limits of aggregation: gains are largest in the realistic middle regime where hardware quality differs meaningfully across clients, and smaller on harder settings such as FMNIST or deeper/noisier configurations where all methods are constrained by overall optimization difficulty. In that sense, Q-RAIL is a practical robustness layer within the QFL workflow.

Several next steps follow naturally. The first is validation on live hardware with time-varying calibrations rather than snapshot-based fake backends. A second is broader cross-platform portability, so the same reliability-aware logic can be tested beyond the current Qiskit/BackendV2 setting. A third is tighter integration with personalization, adversarial robustness, and possibly fairness-aware federation. Another important direction is the theoretical analysis of convergence and weighting behavior under backend-dependent noise. Finally, a particularly promising research avenue is an adaptive Q-RAIL framework in which the stabilization parameters $\tau$, $\beta$, and $w_{min}$ are adjusted online rather than fixed a priori. 

\vspace{-5pt}
\section*{Acknowledgments}
\vspace{-4pt}
This work was supported by the NYUAD Center for Cyber Security (CCS), funded by Tamkeen under the NYUAD Research Institute Award G1104.

\bibliographystyle{IEEEtran}
\bibliography{main}

\end{spacing}

\end{document}

%% file: Tables/transpilation.tex
\begin{table}[t]
\centering
\scriptsize
\setlength{\tabcolsep}{1.25pt}
\renewcommand{\arraystretch}{1.12}

\caption{Transpilation statistics of the same 5-qubit, 12-layer
ring-entangled VQC across IBM fake backends at optimization
levels 1--3.}
\label{tab:backend-transpilation}

\begin{tabular}{|c|c|c|r|r|r|r|c|c|}
\hline
\textbf{Backend} &
\shortstack{\textbf{Opt.}\\\textbf{level}} &
\shortstack{\textbf{2Q}\\\textbf{basis}} &
\shortstack{\textbf{Circuit}\\\textbf{depth}} &
\shortstack{\textbf{1Q}\\\textbf{gates}} &
\shortstack{\textbf{2Q}\\\textbf{gates}} &
\shortstack{\textbf{Routing}\\\textbf{SWAPs}} &
\shortstack{\textbf{Avg. 2Q}\\\textbf{error}} &
\shortstack{\textbf{Avg. RO}\\\textbf{error}} \\
\hline

Logical
& -- & CX & 87 & 130 & 60 & 0 & -- & -- \\
\hline

\multirow{3}{*}{Prague}
& 1 & \multirow{3}{*}{CZ}
& 518 & 929 & 195 & 45
& \multirow{3}{*}{0.002}
& \multirow{3}{*}{0.010} \\
\cline{2-2}\cline{4-7}
& 2 & & 490 & 886 & 187 & 45 & & \\
\cline{2-2}\cline{4-7}
& 3 & & 490 & 886 & 187 & 45 & & \\
\hline

\multirow{3}{*}{Santiago}
& 1 & \multirow{3}{*}{CX}
& 294 & 325 & 252 & 66
& \multirow{3}{*}{0.006}
& \multirow{3}{*}{0.011} \\
\cline{2-2}\cline{4-7}
& 2 & & 340 & 390 & 230 & 64 & & \\
\cline{2-2}\cline{4-7}
& 3 & & 340 & 390 & 230 & 64 & & \\
\hline

\multirow{3}{*}{Cambridge}
& 1 & \multirow{3}{*}{CX}
& 211 & 130 & 199 & 49
& \multirow{3}{*}{0.026}
& \multirow{3}{*}{0.084} \\
\cline{2-2}\cline{4-7}
& 2 & & 209 & 154 & 184 & 46 & & \\
\cline{2-2}\cline{4-7}
& 3 & & 209 & 154 & 184 & 46 & & \\
\hline

\multirow{3}{*}{Manhattan}
& 1 & \multirow{3}{*}{CX}
& 294 & 325 & 252 & 66
& \multirow{3}{*}{0.011}
& \multirow{3}{*}{0.020} \\
\cline{2-2}\cline{4-7}
& 2 & & 305 & 354 & 186 & 46 & & \\
\cline{2-2}\cline{4-7}
& 3 & & 305 & 354 & 186 & 46 & & \\
\hline

\multirow{3}{*}{Torino}
& 1 & \multirow{3}{*}{CZ}
& 539 & 941 & 201 & 47
& \multirow{3}{*}{0.003}
& \multirow{3}{*}{0.023} \\
\cline{2-2}\cline{4-7}
& 2 & & 499 & 892 & 190 & 46 & & \\
\cline{2-2}\cline{4-7}
& 3 & & 499 & 892 & 190 & 46 & & \\
\hline

\multirow{3}{*}{Kyoto}
& 1 & \multirow{3}{*}{ECR}
& 582 & 1242 & 201 & 47
& \multirow{3}{*}{1.000}
& \multirow{3}{*}{0.011} \\
\cline{2-2}\cline{4-7}
& 2 & & 675 & 1208 & 194 & 46 & & \\
\cline{2-2}\cline{4-7}
& 3 & & 674 & 1205 & 194 & 46 & & \\
\hline

\end{tabular}
\vspace{-20pt}
\end{table}

%% file: Algorithms/Algorithms1.tex
\begin{algorithm}[t]
\caption{Proposed Q-RAIL Algorithm}
\scriptsize
\begin{algorithmic}[1]
\Require $\{D_k\}_{k=1}^K$, $B$, $Q$, $\rho$, $T$, $S$, $\theta^0$
\Ensure $\theta^T$, global/local metrics, weights

\State $B_Q \gets \{b \in B : \mathrm{num\_qubits}(b) \ge Q\}$
\ForAll{$b \in B_Q$} 
    \State Compute backend score $C_b$
\EndFor
\State Sort $B_Q$ by $C_b$; let $B_{\mathrm{good}}$ be top-5, $B_{\mathrm{bad}}$ bottom-5, $b^* \gets \arg\min_b C_b$
\State Partition data across $K$ clients \Comment{IID or non-IID}
\State Randomly assign each client $k$ a backend $b_k \in B_{\mathrm{good}} \cup B_{\mathrm{bad}}$ according to $\rho$
\State Instantiate shared VQC $f(x;\theta)$
\ForAll{clients $k$}
    \State Transpile $f$ to $b_k$, extract extract $\{d_k, n_k^{1q}, n_k^{2q}, n_k^{m}\}$, compute $E_k$
\EndFor
\For{$t = 0,\dots,T-1$}
    \For{$k = 1,\dots,K$}
        \State $\theta_k^{t,0} \gets \theta^t$
        \State Run $S$ SPSA steps on $D_k$ using backend $b_k$ to get $\theta_k^{t,\mathrm{loc}}$
        \State $\Delta_k^t \gets \theta_k^{t,\mathrm{loc}} - \theta^t$
    \EndFor
    \State Normalize $\{E_k\}$ to $\{\hat{E}_k\}$ in $[0,1]$
    \State $p_k \gets \dfrac{e^{-\tau \hat{E}_k}}{\sum_j e^{-\tau \hat{E}_j}}$
    \State $\bar{w}_k \gets (1-\beta)p_k + \beta/K$
    \State $w_k^t \gets \dfrac{\max(\bar{w}_k,w_{\min})}{\sum_j \max(\bar{w}_j,w_{\min})}$
    \State $\theta^{t+1} \gets \theta^t + \sum_{k=1}^K w_k^t \Delta_k^t$
    \State Evaluate $\theta^{t+1}$ on $b^*$ and record global metrics
\EndFor
\end{algorithmic}
\label{alg:Alg1}
\end{algorithm}

%% file: Tables/ExpSetup.tex
\begin{table}[t]
\centering
\caption{Experimental design matrix.}
\label{tab:experimental_design}
\scriptsize
\setlength{\tabcolsep}{2pt}
\renewcommand{\arraystretch}{1.18}
\begin{tabular}{|C{0.24\columnwidth}|C{0.10\columnwidth}|C{0.30\columnwidth}|C{0.26\columnwidth}|}
\hline
\textbf{Study} & \textbf{Cfg.} & \textbf{Data / Methods} & \textbf{Controlled setting} \\
\hline
\multicolumn{4}{|c|}{\emph{Benchmark comparisons}} \\
\hline
Main benchmark 
& A 
& \makecell{M / FM / O\\QRAIL /FedAvg /WpQFL}
& IID vs. non-IID \\
\hline
wpQFL-scale check 
& B 
& \makecell{M\\QRAIL /FedAvg /WpQFL}
& $Q=8,\ L=16$ scale \\
\hline
\multicolumn{4}{|c|}{\emph{Robustness and scaling studies}} \\
\hline
Bad-client sweep 
& A 
& \makecell{M, IID\\QRAIL /FedAvg /WpQFL}
& $\rho \in \{0.2,0.5,0.8,1.0\}$ \\
\hline
Client-count sweep 
& A 
& \makecell{M, IID\\QRAIL /FedAvg /WpQFL}
& $K \in \{10,20,50\}$ \\
\hline
Ansatz/CX stress 
& A 
& \makecell{M, IID\\QRAIL / FedAvg}
& \makecell{Linear/Ring/Full\\ CX $\in \{1,3,5,9\}$} \\
\hline
Qubit-scale study 
& -- 
& \makecell{M, IID\\QRAIL / FedAvg}
& $Q \in \{4,10,15\}$ \\
\hline
\hline
\multicolumn{4}{|c|}{\makecell{Shared protocol: angle-encoded PCA inputs, SPSA optimizer, batch size 16, 128/256 shots,\\ 15 rounds, and 5 seeds.}} \\

\hline
\end{tabular}

\vspace{2pt}
\begin{minipage}{0.98\columnwidth}
\scriptsize
\textit{Legend.}
A: $Q=4,L=4,K=10$; B: $Q=8,L=16$.
M: MNIST; FM: Fashion-MNIST; O: OrganAMNIST.
Default $\rho=0.8$ unless swept. Quantum circuit is executed 128 times during training and 256 times during evaluation to estimate measurement probabilities from repeated samples.
\end{minipage}
\label{tab:expsetup}
\vspace{-10pt}
\end{table}

%% file: Tables/table0.tex
\begin{table}
\centering
\scriptsize
\caption{\textbf{Client hardware heterogeneity} (10 clients). Composite error score; lower is better.}
\label{tab:BackendsDistribution}
\begin{tabular}{lcccc}
\toprule
Backend & Group & Qubits & Composite $\downarrow$ & \#Clients \\
\midrule
\texttt{fake\_prague}    & good & 33  & 4.32  & 1 \\
\texttt{fake\_santiago}  & good & 5   & 4.70  & 1 \\
\texttt{fake\_cambridge} & bad  & 28  & 70.11 & 2 \\
\texttt{fake\_manhattan} & bad  & 65  & 99.43 & 2 \\
\texttt{fake\_torino}    & bad  & 133 & 163.79& 3 \\
\texttt{fake\_kyoto}     & bad  & 127 & 301.47& 1 \\
\bottomrule
\end{tabular}
\vspace{-0.35cm}
\end{table}

%% file: Tables/table1.tex
\begin{table*}
\centering
\caption{Final-round results for the main comparison. Mean $\pm$ std across seeds for global train/test loss, accuracy, and AUC, together with average local accuracy and local AUC, for FedAvg, Q-RAIL, and wpQFL under the main comparison settings.}
\setlength{\tabcolsep}{4pt}
\renewcommand{\arraystretch}{1.12}
\resizebox{\textwidth}{!}{%
\begin{tabular}{|c|c|c|l|c|c|c|c|c|c|c|c|}
\hline
\textbf{Config} & \textbf{Dataset Type} & \textbf{Split} & \textbf{Method} & \textbf{Train Loss} & \textbf{Train Acc} & \textbf{Train AUC} & \textbf{Test Loss} & \textbf{Test Acc} & \textbf{Test AUC} & \textbf{Local Acc} & \textbf{Local AUC} \\
\hline
\multirow{24}{*}{A} & \multirow{8}{*}{MNIST} & \multirow{4}{*}{IID} & FedAvg & 0.730 $\pm$ 0.126 & 0.768 $\pm$ 0.072 & 0.916 $\pm$ 0.063 & 0.722 $\pm$ 0.131 & 0.777 $\pm$ 0.084 & 0.920 $\pm$ 0.071 & 0.706 $\pm$ 0.072 & 0.867 $\pm$ 0.058 \\
 &  &  & Q-RAIL & \textbf{0.598 $\pm$ 0.116} & \textbf{0.858 $\pm$ 0.071} & \textbf{0.966 $\pm$ 0.021} & \textbf{0.585 $\pm$ 0.112} & \textbf{0.877 $\pm$ 0.040} & \textbf{0.973 $\pm$ 0.015} & \textbf{0.791 $\pm$ 0.055} & \textbf{0.914 $\pm$ 0.023} \\
 &  &  & wpQFL (euc) & 0.687 $\pm$ 0.111 & 0.814 $\pm$ 0.072 & 0.940 $\pm$ 0.060 & 0.679 $\pm$ 0.115 & 0.833 $\pm$ 0.076 & 0.951 $\pm$ 0.061 & 0.753 $\pm$ 0.066 & 0.894 $\pm$ 0.051 \\
 &  &  & wpQFL (wtd) & 0.684 $\pm$ 0.136 & 0.806 $\pm$ 0.090 & 0.932 $\pm$ 0.064 & 0.680 $\pm$ 0.137 & 0.809 $\pm$ 0.105 & 0.938 $\pm$ 0.069 & 0.747 $\pm$ 0.083 & 0.885 $\pm$ 0.057 \\
\cline{3-12}
 &  & \multirow{4}{*}{non-IID} & FedAvg & 0.789 $\pm$ 0.134 & 0.704 $\pm$ 0.115 & 0.855 $\pm$ 0.084 & 0.774 $\pm$ 0.148 & 0.722 $\pm$ 0.119 & 0.859 $\pm$ 0.099 & 0.809 $\pm$ 0.037 & 0.843 $\pm$ 0.100 \\
 &  &  & Q-RAIL & \textbf{0.680 $\pm$ 0.060} & \textbf{0.797 $\pm$ 0.084} & \textbf{0.943 $\pm$ 0.043} & \textbf{0.655 $\pm$ 0.078} & \textbf{0.813 $\pm$ 0.091} & \textbf{0.952 $\pm$ 0.041} & \textbf{0.851 $\pm$ 0.025} & \textbf{0.897 $\pm$ 0.060} \\
 &  &  & wpQFL (euc) & 0.766 $\pm$ 0.107 & 0.726 $\pm$ 0.122 & 0.883 $\pm$ 0.081 & 0.755 $\pm$ 0.120 & 0.740 $\pm$ 0.130 & 0.891 $\pm$ 0.093 & 0.810 $\pm$ 0.024 & 0.816 $\pm$ 0.142 \\
 &  &  & wpQFL (wtd) & 0.727 $\pm$ 0.114 & 0.732 $\pm$ 0.118 & 0.894 $\pm$ 0.068 & 0.717 $\pm$ 0.123 & 0.760 $\pm$ 0.120 & 0.902 $\pm$ 0.080 & 0.839 $\pm$ 0.045 & 0.834 $\pm$ 0.134 \\
\cline{2-12}
 & \multirow{8}{*}{FMNIST} & \multirow{4}{*}{IID} & FedAvg & \textbf{0.815 $\pm$ 0.036} & 0.646 $\pm$ 0.050 & 0.824 $\pm$ 0.026 & \textbf{0.817 $\pm$ 0.031} & \textbf{0.654 $\pm$ 0.049} & 0.827 $\pm$ 0.026 & \textbf{0.595 $\pm$ 0.046} & 0.779 $\pm$ 0.032 \\
 &  &  & Q-RAIL & 0.830 $\pm$ 0.092 & 0.631 $\pm$ 0.137 & \textbf{0.834 $\pm$ 0.069} & 0.833 $\pm$ 0.073 & 0.626 $\pm$ 0.112 & \textbf{0.832 $\pm$ 0.060} & 0.592 $\pm$ 0.102 & \textbf{0.787 $\pm$ 0.055} \\
 &  &  & wpQFL (euc) & 0.821 $\pm$ 0.020 & 0.631 $\pm$ 0.060 & 0.817 $\pm$ 0.023 & 0.829 $\pm$ 0.015 & 0.625 $\pm$ 0.055 & 0.814 $\pm$ 0.023 & 0.579 $\pm$ 0.026 & 0.770 $\pm$ 0.015 \\
 &  &  & wpQFL (wtd) & 0.826 $\pm$ 0.053 & \textbf{0.647 $\pm$ 0.070} & 0.822 $\pm$ 0.044 & 0.828 $\pm$ 0.039 & 0.645 $\pm$ 0.064 & 0.822 $\pm$ 0.034 & 0.590 $\pm$ 0.062 & 0.776 $\pm$ 0.043 \\
\cline{3-12}
 &  & \multirow{4}{*}{non-IID} & FedAvg & 0.857 $\pm$ 0.102 & 0.635 $\pm$ 0.117 & 0.827 $\pm$ 0.080 & 0.862 $\pm$ 0.090 & 0.640 $\pm$ 0.098 & 0.823 $\pm$ 0.071 & 0.690 $\pm$ 0.097 & 0.826 $\pm$ 0.056 \\
 &  &  & Q-RAIL & \textbf{0.853 $\pm$ 0.113} & \textbf{0.651 $\pm$ 0.104} & \textbf{0.831 $\pm$ 0.076} & \textbf{0.857 $\pm$ 0.099} & 0.631 $\pm$ 0.087 & \textbf{0.827 $\pm$ 0.062} & \textbf{0.710 $\pm$ 0.068} & 0.827 $\pm$ 0.079 \\
 &  &  & wpQFL (euc) & 0.877 $\pm$ 0.092 & 0.611 $\pm$ 0.084 & 0.796 $\pm$ 0.074 & 0.886 $\pm$ 0.076 & 0.596 $\pm$ 0.071 & 0.789 $\pm$ 0.063 & 0.664 $\pm$ 0.071 & 0.800 $\pm$ 0.049 \\
 &  &  & wpQFL (wtd) & \textbf{0.853 $\pm$ 0.082} & 0.645 $\pm$ 0.091 & 0.830 $\pm$ 0.066 & 0.859 $\pm$ 0.064 & \textbf{0.645 $\pm$ 0.058} & 0.826 $\pm$ 0.051 & 0.691 $\pm$ 0.081 & \textbf{0.836 $\pm$ 0.055} \\
\cline{2-12}
 & \multirow{8}{*}{OrganAMNIST} & \multirow{4}{*}{IID} & FedAvg & 0.839 $\pm$ 0.034 & 0.668 $\pm$ 0.062 & 0.857 $\pm$ 0.030 & 0.874 $\pm$ 0.037 & 0.650 $\pm$ 0.051 & 0.838 $\pm$ 0.021 & 0.614 $\pm$ 0.029 & 0.810 $\pm$ 0.022 \\
 &  &  & Q-RAIL & \textbf{0.815 $\pm$ 0.036} & \textbf{0.735 $\pm$ 0.033} & \textbf{0.866 $\pm$ 0.013} & \textbf{0.851 $\pm$ 0.025} & \textbf{0.704 $\pm$ 0.018} & \textbf{0.846 $\pm$ 0.011} & \textbf{0.656 $\pm$ 0.029} & 0.814 $\pm$ 0.021 \\
 &  &  & wpQFL (euc) & 0.849 $\pm$ 0.029 & 0.655 $\pm$ 0.029 & 0.857 $\pm$ 0.022 & 0.886 $\pm$ 0.041 & 0.628 $\pm$ 0.029 & 0.833 $\pm$ 0.021 & 0.602 $\pm$ 0.021 & 0.803 $\pm$ 0.016 \\
 &  &  & wpQFL (wtd) & 0.818 $\pm$ 0.058 & 0.694 $\pm$ 0.078 & 0.862 $\pm$ 0.012 & 0.855 $\pm$ 0.059 & 0.661 $\pm$ 0.065 & 0.843 $\pm$ 0.019 & 0.639 $\pm$ 0.068 & \textbf{0.818 $\pm$ 0.020} \\
\cline{3-12}
 &  & \multirow{4}{*}{non-IID} & FedAvg & 0.872 $\pm$ 0.097 & 0.660 $\pm$ 0.066 & 0.820 $\pm$ 0.062 & 0.904 $\pm$ 0.069 & 0.605 $\pm$ 0.048 & 0.809 $\pm$ 0.045 & 0.661 $\pm$ 0.109 & \textbf{0.775 $\pm$ 0.083} \\
 &  &  & Q-RAIL & \textbf{0.858 $\pm$ 0.069} & \textbf{0.670 $\pm$ 0.056} & \textbf{0.830 $\pm$ 0.045} & \textbf{0.895 $\pm$ 0.056} & \textbf{0.621 $\pm$ 0.046} & \textbf{0.810 $\pm$ 0.039} & \textbf{0.674 $\pm$ 0.088} & 0.772 $\pm$ 0.059 \\
 &  &  & wpQFL (euc) & 0.911 $\pm$ 0.091 & 0.632 $\pm$ 0.037 & 0.787 $\pm$ 0.063 & 0.922 $\pm$ 0.072 & 0.599 $\pm$ 0.031 & 0.789 $\pm$ 0.048 & 0.635 $\pm$ 0.092 & 0.734 $\pm$ 0.119 \\
 &  &  & wpQFL (wtd) & 0.874 $\pm$ 0.082 & 0.663 $\pm$ 0.062 & 0.824 $\pm$ 0.053 & 0.905 $\pm$ 0.063 & 0.617 $\pm$ 0.064 & 0.809 $\pm$ 0.036 & 0.640 $\pm$ 0.106 & 0.768 $\pm$ 0.081 \\
\hline
\multirow{8}{*}{B} & \multirow{8}{*}{MNIST} & \multirow{4}{*}{IID} & FedAvg & 0.937 $\pm$ 0.100 & \textbf{0.659 $\pm$ 0.124} & 0.794 $\pm$ 0.152 & 0.944 $\pm$ 0.103 & \textbf{0.641 $\pm$ 0.116} & 0.798 $\pm$ 0.160 & 0.513 $\pm$ 0.078 & 0.699 $\pm$ 0.076 \\
 &  &  & Q-RAIL & \textbf{0.926 $\pm$ 0.118} & 0.643 $\pm$ 0.148 & \textbf{0.805 $\pm$ 0.151} & \textbf{0.930 $\pm$ 0.122} & 0.627 $\pm$ 0.152 & 0.810 $\pm$ 0.164 & 0.513 $\pm$ 0.097 & 0.705 $\pm$ 0.088 \\
 &  &  & wpQFL (euc) & 0.956 $\pm$ 0.086 & 0.605 $\pm$ 0.055 & 0.780 $\pm$ 0.141 & 0.965 $\pm$ 0.083 & 0.588 $\pm$ 0.036 & 0.783 $\pm$ 0.143 & 0.502 $\pm$ 0.059 & 0.690 $\pm$ 0.068 \\
 &  &  & wpQFL (wtd) & 0.935 $\pm$ 0.099 & 0.652 $\pm$ 0.114 & \textbf{0.805 $\pm$ 0.145} & 0.941 $\pm$ 0.101 & 0.629 $\pm$ 0.104 & \textbf{0.812 $\pm$ 0.155} & \textbf{0.522 $\pm$ 0.074} & \textbf{0.710 $\pm$ 0.073} \\
\cline{3-12}
 &  & \multirow{4}{*}{non-IID} & FedAvg & 0.998 $\pm$ 0.069 & \textbf{0.582 $\pm$ 0.063} & 0.739 $\pm$ 0.138 & 1.010 $\pm$ 0.068 & \textbf{0.564 $\pm$ 0.053} & 0.744 $\pm$ 0.140 & \textbf{0.524 $\pm$ 0.113} & 0.651 $\pm$ 0.177 \\
 &  &  & Q-RAIL & \textbf{0.986 $\pm$ 0.072} & 0.532 $\pm$ 0.111 & \textbf{0.747 $\pm$ 0.104} & \textbf{0.986 $\pm$ 0.087} & 0.519 $\pm$ 0.140 & \textbf{0.749 $\pm$ 0.114} & 0.502 $\pm$ 0.180 & 0.609 $\pm$ 0.158 \\
 &  &  & wpQFL (euc) & 1.022 $\pm$ 0.087 & 0.574 $\pm$ 0.052 & 0.704 $\pm$ 0.140 & 1.037 $\pm$ 0.084 & 0.554 $\pm$ 0.040 & 0.705 $\pm$ 0.141 & 0.503 $\pm$ 0.117 & \textbf{0.665 $\pm$ 0.077} \\
 &  &  & wpQFL (wtd) & 1.004 $\pm$ 0.080 & 0.562 $\pm$ 0.098 & 0.731 $\pm$ 0.153 & 1.015 $\pm$ 0.075 & 0.546 $\pm$ 0.079 & 0.737 $\pm$ 0.150 & 0.515 $\pm$ 0.142 & 0.646 $\pm$ 0.179 \\
\hline
\end{tabular}%
}
\label{tab:mainComparison}
\end{table*}

%% file: Tables/table2.tex
\begin{table*}
\centering
\caption{Detailed bad-client sweep results. Mean ± std across seeds for final global metrics and bad-client local metrics under different bad-client proportions. Bad-client local metrics are averaged first across bad clients within each seed, then across seeds.}
\setlength{\tabcolsep}{4pt}
\renewcommand{\arraystretch}{1.12}
\resizebox{\textwidth}{!}{%
\begin{tabular}{|c|l|c|c|c|c|c|c|c|c|c|}
\hline
\textbf{Bad clients} & \textbf{Method} & \textbf{Train Loss} & \textbf{Train Acc} & \textbf{Train AUC} & \textbf{Test Loss} & \textbf{Test Acc} & \textbf{Test AUC} & \textbf{Bad Local Loss} & \textbf{Bad Local Acc} & \textbf{Bad Local AUC} \\
\hline
\multirow{4}{*}{20\%}
& FedAvg          & 0.695 $\pm$ 0.110 & 0.815 $\pm$ 0.069 & 0.950 $\pm$ 0.031 & 0.685 $\pm$ 0.111 & 0.813 $\pm$ 0.070 & 0.959 $\pm$ 0.031 & 0.787 $\pm$ 0.098 & 0.784 $\pm$ 0.059 & 0.942 $\pm$ 0.037 \\
& Q-RAIL      & \textbf{0.655 $\pm$ 0.092} & 0.846 $\pm$ 0.082 & \textbf{0.957 $\pm$ 0.037} & \textbf{0.641 $\pm$ 0.093} & \textbf{0.851 $\pm$ 0.089} & \textbf{0.967 $\pm$ 0.033} & 0.759 $\pm$ 0.094 & \textbf{0.804 $\pm$ 0.083} & \textbf{0.948 $\pm$ 0.046} \\
& wpQFL-euclidean & 0.675 $\pm$ 0.136 & \textbf{0.852 $\pm$ 0.098} & 0.950 $\pm$ 0.056 & 0.663 $\pm$ 0.147 & 0.847 $\pm$ 0.114 & 0.952 $\pm$ 0.065 & 0.766 $\pm$ 0.110 & 0.796 $\pm$ 0.098 & 0.941 $\pm$ 0.061 \\
& wpQFL-weighted  & 0.658 $\pm$ 0.111 & 0.827 $\pm$ 0.083 & 0.956 $\pm$ 0.034 & 0.655 $\pm$ 0.119 & 0.827 $\pm$ 0.083 & 0.962 $\pm$ 0.034 & \textbf{0.757 $\pm$ 0.095} & 0.791 $\pm$ 0.079 & \textbf{0.948 $\pm$ 0.041} \\
\hline
\multirow{4}{*}{50\%}
& FedAvg          & \textbf{0.618 $\pm$ 0.096} & 0.842 $\pm$ 0.049 & 0.959 $\pm$ 0.030 & \textbf{0.614 $\pm$ 0.104} & 0.845 $\pm$ 0.068 & 0.965 $\pm$ 0.030 & \textbf{0.726 $\pm$ 0.082} & 0.813 $\pm$ 0.058 & 0.947 $\pm$ 0.045 \\
& Q-RAIL      & 0.629 $\pm$ 0.119 & 0.836 $\pm$ 0.074 & 0.947 $\pm$ 0.045 & 0.620 $\pm$ 0.126 & 0.850 $\pm$ 0.079 & 0.952 $\pm$ 0.051 & \textbf{0.726 $\pm$ 0.093} & 0.818 $\pm$ 0.075 & 0.936 $\pm$ 0.057 \\
& wpQFL-euclidean & 0.659 $\pm$ 0.060 & \textbf{0.843 $\pm$ 0.054} & \textbf{0.968 $\pm$ 0.015} & 0.647 $\pm$ 0.060 & \textbf{0.858 $\pm$ 0.047} & \textbf{0.977 $\pm$ 0.011} & 0.748 $\pm$ 0.067 & \textbf{0.829 $\pm$ 0.044} & \textbf{0.961 $\pm$ 0.019} \\
& wpQFL-weighted  & 0.653 $\pm$ 0.100 & 0.841 $\pm$ 0.055 & 0.958 $\pm$ 0.030 & 0.645 $\pm$ 0.101 & 0.846 $\pm$ 0.069 & 0.963 $\pm$ 0.034 & 0.749 $\pm$ 0.089 & 0.818 $\pm$ 0.055 & 0.952 $\pm$ 0.032 \\
\hline
\multirow{4}{*}{80\%}
& FedAvg          & 0.730 $\pm$ 0.126 & 0.768 $\pm$ 0.072 & 0.916 $\pm$ 0.063 & 0.722 $\pm$ 0.131 & 0.777 $\pm$ 0.084 & 0.920 $\pm$ 0.071 & 0.847 $\pm$ 0.100 & 0.684 $\pm$ 0.071 & 0.854 $\pm$ 0.058 \\
& Q-RAIL      & \textbf{0.598 $\pm$ 0.116} & \textbf{0.858 $\pm$ 0.071} & \textbf{0.966 $\pm$ 0.021} & \textbf{0.585 $\pm$ 0.112} & \textbf{0.877 $\pm$ 0.040} & \textbf{0.973 $\pm$ 0.015} & \textbf{0.760 $\pm$ 0.086} & \textbf{0.772 $\pm$ 0.051} & \textbf{0.902 $\pm$ 0.022} \\
& wpQFL-euclidean & 0.687 $\pm$ 0.111 & 0.814 $\pm$ 0.072 & 0.940 $\pm$ 0.060 & 0.679 $\pm$ 0.115 & 0.833 $\pm$ 0.076 & 0.951 $\pm$ 0.061 & 0.813 $\pm$ 0.084 & 0.731 $\pm$ 0.060 & 0.880 $\pm$ 0.052 \\
& wpQFL-weighted  & 0.684 $\pm$ 0.136 & 0.806 $\pm$ 0.090 & 0.932 $\pm$ 0.064 & 0.680 $\pm$ 0.137 & 0.809 $\pm$ 0.105 & 0.938 $\pm$ 0.069 & 0.815 $\pm$ 0.094 & 0.727 $\pm$ 0.078 & 0.872 $\pm$ 0.056 \\
\hline
\multirow{4}{*}{100\%}
& FedAvg          & 0.756 $\pm$ 0.125 & 0.784 $\pm$ 0.106 & 0.900 $\pm$ 0.070 & 0.740 $\pm$ 0.140 & 0.783 $\pm$ 0.135 & 0.909 $\pm$ 0.081 & 0.857 $\pm$ 0.096 & 0.706 $\pm$ 0.121 & 0.851 $\pm$ 0.076 \\
& Q-RAIL      & \textbf{0.747 $\pm$ 0.135} & \textbf{0.785 $\pm$ 0.104} & \textbf{0.914 $\pm$ 0.057} & \textbf{0.730 $\pm$ 0.151} & \textbf{0.794 $\pm$ 0.106} & \textbf{0.925 $\pm$ 0.062} & \textbf{0.849 $\pm$ 0.105} & \textbf{0.712 $\pm$ 0.085} & \textbf{0.861 $\pm$ 0.060} \\
& wpQFL-euclidean & 0.770 $\pm$ 0.140 & 0.752 $\pm$ 0.144 & 0.908 $\pm$ 0.078 & 0.755 $\pm$ 0.151 & 0.754 $\pm$ 0.152 & 0.918 $\pm$ 0.082 & 0.866 $\pm$ 0.109 & 0.683 $\pm$ 0.126 & 0.855 $\pm$ 0.080 \\
& wpQFL-weighted  & 0.759 $\pm$ 0.144 & 0.781 $\pm$ 0.137 & 0.894 $\pm$ 0.088 & 0.747 $\pm$ 0.158 & 0.793 $\pm$ 0.144 & 0.898 $\pm$ 0.098 & 0.861 $\pm$ 0.108 & 0.705 $\pm$ 0.129 & 0.847 $\pm$ 0.087 \\
\hline
\end{tabular}%
}
\label{tab:BadClientsSweep}
\end{table*}